\documentclass[showpacs,amsmath,amssymb, prb,twocolumn]{revtex4}
\usepackage{graphicx}
\usepackage{dcolumn}
\usepackage{bm}
\begin{document}

\title{A topological current divider}

\author{Francesco Romeo}
\affiliation{ Dipartimento di Fisica ''E. R. Caianiello'',
	Universit\`a degli
	Studi di Salerno, Via Giovanni Paolo II, I-84084 Fisciano (Sa), Italy}

\begin{abstract}
We study the transport properties of a hybrid junction made of a ferromagnetic lead in electrical connection with the helical edge modes of a two-dimensional topological insulator. In this system, the time reversal symmetry, which characterizes the ballistic edge modes of the topological insulator, is explicitly broken inside the ferromagnetic region. This conflict situation generates unusual transport phenomena at the interface which are the  manifestation of the interplay between the spin polarization of the injected current and the spin-momentum locking mechanism operating inside the topological insulator. We show that the spin polarized current originated in the ferromagnetic region is asymmetrically divided in spatially separated branch currents sustained by edge channels with different helicity inside the topological insulator. The above findings provide the working principle of a topological current divider in which the relative intensity of the branch currents is determined by the polarization of the incoming current. We discuss the relevance of this effect in spintronics where, for instance, it offers an alternative way to measure the current polarization generated by a ferromagnetic electrode.
\end{abstract}
\maketitle
\section{Introduction}
\label{sec:intro}
The search for new topological states of matter is one of the most active field in physics \cite{topo1,topo2,topo3,topo4,topo5,topo6,topo7}. The scientific interest towards topological states is motivated by their robustness against material defects and imperfections, which is a desirable property for technological applications and fundamental studies. The robustness of topological matter against perturbations originates from the existence of boundary modes which are protected by the symmetries of the bulk. These topological modes can be described by using low-energy models in reduced spatial dimension, the resulting theories being representative of the whole material. This intriguing situation shares analogies with the holographic principle \cite{holo} which is a supposed property of quantum gravity inspired by the black hole thermodynamics.\\
Topological superconductors \cite{topsuper} and topological insulators \cite{topins} are important members of the topological matter family. The research lines focused on these states of matter are not completely independent since topological superconductivity can be achieved, for instance, by proximizing a topological insulator with a conventional s-wave superconductor\cite{topsupins}.\\
Since the first theoretical proposal by Kitaev \cite{kitaev}, the interest for topological superconductivity has been fueled by the search for the Majorana's quasiparticle \cite{majoexp1,majoexp2}. The latter has produced an intense theoretical\cite{theormajo,theormajo2} and experimental\cite{expmajo1,expmajo2,expmajo3} activity inspired by the possibility to obtain useful information for the forthcoming quantum computers \cite{kitaev-computation,majo-computation}.\\
On the other hand, topological insulator state in two-dimension has been first synthesized in mercury-telluride quantum wells more than ten years ago\cite{2dTiQW1,2dTiQW2}. Scientific efforts in this direction lead to the theorization\cite{3dTith} and subsequent discovery\cite{3dTi} of three-dimensional topological insulators. Recently, the more exotic category of high-order topological insulators\cite{hoti1,hoti2,hoti3,hoti4} hosting protected surface, edge or corner states has been introduced. Two-dimensional topological insulators are particularly appealing for applications and fundamental studies\cite{dolcini1,dolcini2,dolcini3,dolcini4,dolcini5,dolcini6,dolcini7,romeo1,romeo2,romeo3,romeo4,bercioux1}. These systems present a bulk insulating phase accompanied by one dimensional conducting edge modes with preserved helicity. These states are protected from backscattering effects by the time reversal symmetry and behave like ideal ballistic channels presenting the spin-momentum locking effect, which is very appealing in spintronics\cite{ABferroTi,bercioux1,F-TIdev1,F-TIdev2,F-TIdev3}. Moreover they are characterized by a linear dispersion relation spanning an energy range inside the bulk gap.\\
The size of the bulk gap is an important material parameter. A large gap is a desirable property for device applications since it ensures that the topological phase is not contaminated by thermally excited non-topological states. Recently, two-dimensional topological insulators with bulk gap of 100 meV have been identified\cite{largegapexp}. This value, which is sensibly greater than the gap values of the HgTe/CdTe and InAs/GaSb quantum wells, is compatible with room temperature applications. These recent developments suggest that a topological protected room-temperature electronics could be soon achieved\cite{QSHEroom}. In view of this revolution, new paradigms are needed to fully exploit the potential of the topological phase.\\
In searching for new effects, an inspiring paradigm is combining systems with heterogeneous characteristics. In these heterostructures, the competition between different orders sometimes generates emerging properties. Inspired by these arguments, in this work we study the transport properties of an hybrid system obtained by coupling the massive states of a ferromagnetic electrode with the massless edge modes of a two-dimensional topological insulator. In this system the time reversal symmetry is explicitly broken inside the ferromagnet, while it is preserved inside the topological insulator. Under this conflict condition, the system response is strongly affected by the influence of magnetism on the helical states\cite{magnetismhelicalexp}. In order to study the system response, we build a low-energy formulation of the problem allowing to study the transport properties of the heterojunction within the framework of the scattering field theory \`{a} la B\"{u}ttiker \cite{buttiker92,blanter2000}. We demonstrate that the application of a voltage bias to the system induces an asymmetric splitting of the polarized current coming from the ferromagnet into two branch currents having relative intensity controlled by the polarization of the incoming current. We discuss the relevance of this effect in spintronics along with the working principle of a topological current divider. Implementation details related to the use of Stoner or spin bandwidth asymmetry ferromagents are also commented.\\
The work is organized as follows. In Sec. \ref{sec:hamMod}, we introduce the problem and provide a low energy effective model for the device. Boundary conditions imposed by the modes hybridization are carefully discussed within the framework of the matching matrix formalism. The scattering matrix is also derived. In Sec. \ref{sec:scatt}, we present the scattering field theory and derive the observables of the system. Commented results are presented in Sec. \ref{sec:results}, while the conclusions are given in Sec. \ref{sec:concl}. Details on the scattering approach are reported in Appendix \ref{sec:MtoS} and \ref{sec:ExpS}.

\begin{figure}[!h]
\includegraphics[scale=0.45]{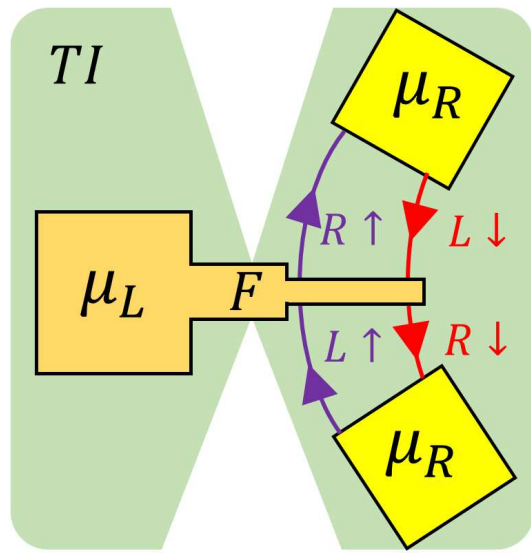}
\caption{Schematic of the topological current divider described in the main text. A two-dimensional topological insulator ($TI$) is laterally etched to form a constriction. A ferromagnetic electrode ($F$) is created in the constriction region where massive electronic states belonging to the $F$ region hybridize with the massless helical modes of the topological insulator.  The application of a voltage bias to the system induces an electrochemical potential gradient ($\mu_{L} \neq \mu_{R}$) responsible for a current. The current coming from the ferromagnetic region is splitted in two branch currents: The top current sustained by edge modes with positive helicity ($R\uparrow$- and $L\downarrow$-states); the bottom current sustained by edge modes with negative helicity ($R\downarrow$- and $L\uparrow$-states). The top and bottom currents have different intensity, this difference being controlled by the current polarization instead of the electrical resistances of the branches.}
\label{figure1}
\end{figure}

\section{Hamiltonian model}
\label{sec:hamMod}
Let us consider the system depicted in Fig. \ref{figure1} consisting of a ferromagnetic electrode in electrical connection with the edge states of a two-dimensional topological insulator. In order to force the modes coupling between the ferromagnetic and the topological region, a constriction is formed by lateral etching of the topological insulator. Adopting a one dimensional description, the Hamiltonian model can be written in the form $H=H_{TI}+H_{F}+H_{t}$, where $H_{TI}$ and $H_{F}$ represent, respectively, the topological insulator and the ferromagnet Hamiltonian. The additional term $H_{t}$ represents the tunneling Hamiltonian, which is left undetermined for the moment. The edge modes of the topological region provide the relevant degrees of freedom contributing to the Hamiltonian and accordingly we can write \cite{chamon}:
\begin{eqnarray}
\label{eq:HamTi}
H_{TI}=-i\hbar v \sum_{\sigma}\int dx \Big (\psi^{\dag}_{R\sigma}\partial_x \psi_{R\sigma}-\psi^{\dag}_{L\sigma}\partial_x \psi_{L\sigma}\Big ),
\end{eqnarray}
where $v$ is the propagation velocity of the edge modes, $\sigma \in \{\uparrow, \downarrow\}$ represents the spin projection, while $\psi_{R\sigma}$ and $\psi_{L\sigma}$ represent fermionic fields in second quantization obeying anticommutation relations. The edge modes with positive helicity ($R\uparrow$- and $L\downarrow$-states) are spatially separated from the edge modes with negative helicity ($R\downarrow$- and $L\uparrow$-states). On the other hand, the Hamiltonian of the ferromagnetic side of the junction is given by
\begin{eqnarray}
\label{eq:HamF}
H_{F}=\sum_{\sigma}\int dx \Bigl [\psi^{\dag}_{\sigma} \Big(-\frac{\hbar^{2}\partial^{2}_x}{2 m_{\sigma}}-\sigma h_{ex} -E_F\Bigl)\psi_{\sigma} \Bigl],
\end{eqnarray}
where have introduced the Fermi energy $E_F$ and the second quantization fermionic fields $\psi_{\sigma}$ which describe massive states with spin projection $\sigma$ and effective mass $m_{\sigma}$. An exchange term $-\sigma h_{ex}$ is included in order to simultaneously account for Stoner and spin bandwidth asymmetry ferromagnetism \cite{hirsch,Higashiguchi}. The quantum states of the topological and ferromagnetic side are hybridized when the tunneling Hamiltonian is considered. The latter statement can be easily verified by writing the Heisenberg equations of the motion for the fermionic fields involved in the problem. Accordingly, we obtain the following equations:
\begin{eqnarray}
\label{eq:EOM}
& & i \hbar \partial_{t} \psi_{R \sigma}=[\psi_{R \sigma},H]=-i\hbar v \partial_{x}\psi_{R \sigma}+[\psi_{R \sigma},H_{t}]\nonumber\\
& & i \hbar \partial_{t} \psi_{L \sigma}=[\psi_{L \sigma},H]=i\hbar v \partial_{x}\psi_{L \sigma}+[\psi_{L \sigma},H_{t}]\\
& & i \hbar \partial_{t} \psi_{\sigma}=[\psi_{\sigma},H]=\Big(-\frac{\hbar^{2}\partial^{2}_x}{2 m_{\sigma}}-\sigma h_{ex}-E_F \Bigl)\psi_{\sigma}+[\psi_{\sigma},H_{t}],\nonumber
\end{eqnarray}
where we have introduced the notation $[\psi,H]=\psi H-H \psi$ meaning the commutator of $\psi$ with the Hamiltonian $H$. Due to the lack of two-body interaction effects, Equations (\ref{eq:EOM}) present the same structure of the Schr\"{o}dinger problem in first quantization. The structure of the problem is completely determined once the coupling terms $[\psi_{R/L \sigma},H_{t}]$ and $[\psi_{\sigma},H_{t}]$ have been assigned by making a specific choice for the tunneling Hamiltonian. Within the framework of a continuous model, the aforementioned coupling terms are sensibly different from zero only at the interface point between the ferromagnetic and the topological region. In the following, without loss of generality, we fix the interface position at $x=0$. Due to the above considerations, the coupling terms originated by the tunneling Hamiltonian can be arranged in the form of an interface potential whose effects can be accounted by using appropriate boundary conditions at the interface.\\
We are interested in describing the transport properties in linear response regime and thus it is convenient to resort to a low-energy projection of the ferromagnet Hamiltonian. The projection provides an accurate description of electronic states with energy eigenvalue close to the Fermi level, i.e. the quantum states which are relevant in defining the transport properties of the system. In the ferromagnetic side of the system ($x<0$) the Schr\"{o}dinger problem can be written in spinorial form as follows:
\begin{eqnarray}
\label{eq:schrodingerF}
\left[
  \begin{array}{cc}
    -\frac{\hbar^{2}\partial^{2}_x}{2 m_{\uparrow}}-h_{ex}-E_F & 0 \\
    0 & -\frac{\hbar^{2}\partial^{2}_x}{2 m_{\downarrow}}+h_{ex}-E_F \\
  \end{array}
\right]\psi=i \hbar \partial_{t}\psi,
\end{eqnarray}
where $\psi=(\psi_\uparrow,\psi_\downarrow)^t$. Following the low-energy projection procedure described in Ref. [\onlinecite{projection}], the wave function in vicinity of the Fermi energy $E_F$ can be expanded by introducing left and right movers representation. In this way we can write:
\begin{eqnarray}
\label{eq:lrmovers}
\psi_{\sigma}=\psi_{R\sigma}(x) e^{ik^{\sigma}_F x}+\psi_{L\sigma}(x) e^{-ik^{\sigma}_F x},
\end{eqnarray}
where $\hbar k^{\sigma}_F=\sqrt{2 m_{\sigma}(E_{F}+\sigma h_{ex})}$ are the spin-sensitive Fermi momenta, while $\psi_{R\sigma}(x)$ and $\psi_{L\sigma}(x)$ are slowly varying functions of the spatial coordinate $x$. Using Eq. (\ref{eq:lrmovers}) in Eq. (\ref{eq:schrodingerF}) and neglecting second derivatives of $\psi_{L/R\sigma}(x)$ and rapidly oscillating terms, the initial Schr\"{o}dinger problem can be written in the form:
\begin{eqnarray}
\label{eq:Fproj}
\left[
  \begin{array}{cc}
    D_{+} & O_{2\times2} \\
    O_{2\times2} & D_{-} \\
  \end{array}
\right] \Psi=i \hbar \partial_t \Psi,
\end{eqnarray}
where $\Psi=(\psi_{R\uparrow},\psi_{L\downarrow},\psi_{R\downarrow},\psi_{L\uparrow})^t$,  $O_{2\times2}$ represents a $2 \times 2$ matrix with vanishing elements, while
\begin{eqnarray}
\label{eq:dmatrix}
D_{\pm}=\left[
          \begin{array}{cc}
            -i\hbar v^{\pm}_{F}\partial_x & 0 \\
            0 & i\hbar v^{\mp}_{F}\partial_x \\
          \end{array}
        \right]
\end{eqnarray}
includes the kinetic energy of left and right movers written in terms of the spin sensitive velocities $v^{\sigma}_{F}=\hbar k^{\sigma}_F/m_{\sigma}$. Within the above representation, which is valid for $x<0$, the relevant information on the ferromagnetic state is provided by the spin-dependent quantities $v^{\sigma}_{F}$, which on their turn depend on the position of the Fermi level. Moreover, in the limit $v^{\sigma}_{F}\rightarrow v$, Eq. (\ref{eq:Fproj}) takes the same form of the Hamiltonian problem of the topological side of the system ($x>0$), i.e.
\begin{eqnarray}
\label{eq:Tiproj}
\left[
  \begin{array}{cc}
    D & O_{2\times2} \\
    O_{2\times2} & D \\
  \end{array}
\right] \Psi=i \hbar \partial_t \Psi,
\end{eqnarray}
with $D=-i\hbar v \partial_{x} \sigma_z$ and $\sigma_z$ the Pauli matrix. In this way, both the sides of the junction can be described by adopting the same spinorial representation. The relation between the wave functions at the interface depends on the tunneling Hamiltonian and can be described by using the matching matrix formalism. In particular, once a tunneling Hamiltonian $H_t$ has been specified, a matching matrix $\mathcal{M}$ is identified. The wave functions in close vicinity of the interface point $x=0$ obey the relation:
\begin{eqnarray}
\label{eq:matching}
\Psi(0^{+})=\mathcal{M}\Psi(0^{-}),
\end{eqnarray}
where $\Psi(0^{+})$ and $\Psi(0^{-})$ represent the wave functions belonging to the topological or ferromagnetic side, respectively. In writing Eq. (\ref{eq:matching}), the notation $x_0^{\pm}=x_0 \pm \epsilon$ has been introduced with $\epsilon$ a positive infinitesimal quantity.\\
So far we have identified a low-energy model which allows the description of the whole system within the same spinorial representation. In the derivation, we have only invoked general properties of the tunneling Hamiltonian $H_t$, with special emphasis on its local character. For presentation reasons, we postpone to the next Section the discussion about the identification of the interface potential generated by $H_t$.

\section{Scattering theory and observables}
\label{sec:scatt}
In the following, we provide a detailed description of the scattering field theory. Boundary conditions for the scattering problem along with a derivation of the $\mathcal{M}$-matrix are also discussed.

\subsection{Scattering fields theory}
\begin{figure}[!h]
\includegraphics[scale=0.55]{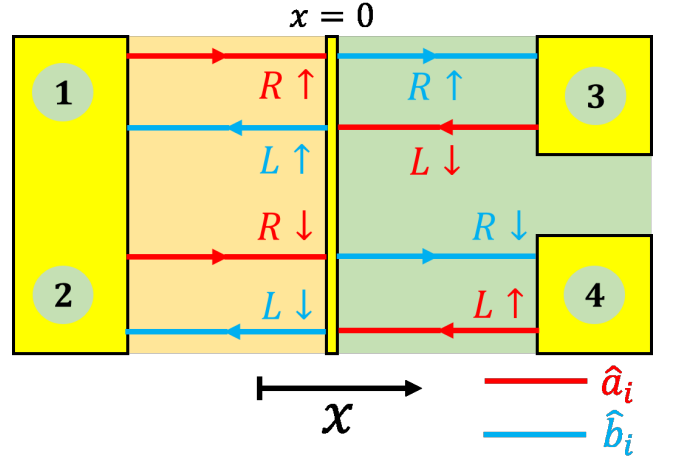}
\caption{Schematic of the device effective model. Modes labeled by $1$ and $2$ belong to the ferromagnetic electrode and represent the spin up and spin down channels, respectively. Modes labeled by $3$ and $4$ belong to the topological insulator and represent positive and negative helicity edge states, respectively.}
\label{figure2}
\end{figure}

Within the framework of the B\"{u}ttiker approach, second quantization scattering fields are introduced. These fields are written in terms of translational invariant eigenmodes of the local Hamiltonian describing each semi-infinite electrode. This construction is quite general and can be adapted to the present situation.\\
Following the above procedure, the scattering field describing elementary processes inside the ferromagnetic lead can be written as follows:
\begin{eqnarray}
\label{eq:fieldF}
\hat{\psi}_F(x,t)&=&\int dE \Big ( \phi_{R\uparrow}(x,t)|1\rangle+\phi_{L\downarrow}(x,t)|2\rangle+\nonumber\\
&+&\phi_{R\downarrow}(x,t)|3\rangle+\phi_{L\uparrow}(x,t)|4\rangle\Big),
\end{eqnarray}
where the notation
\begin{eqnarray}
\label{eq:fieldF2}
&&\phi_{R\sigma}(x,t)=\frac{e^{-iEt/\hbar}}{\sqrt{2 \pi \hbar v^{\sigma}_F}}e^{ik^{\sigma}_{E}x}\hat{a}_{R \sigma}(E)\\
\label{eq:fieldF3}
&&\phi_{L\sigma}(x,t)=\frac{e^{-iEt/\hbar}}{\sqrt{2 \pi \hbar v^{\sigma}_F}}e^{-ik^{\sigma}_{E}x}\hat{b}_{L \sigma}(E)\\
&&k^{\sigma}_{E}=\frac{E}{\hbar v^{\sigma}_F}>0
\end{eqnarray}
has been introduced. Moreover, to implement a multichannel theory, the auxiliary quantities $|1\rangle=(1,0,0,0)^t$, $|2\rangle=(0,1,0,0)^t$, $|3\rangle=(0,0,1,0)^t$, $|4\rangle=(0,0,0,1)^t$ have been defined. The scattering operators $\hat{a}_{R \sigma}(E)$ and $\hat{b}_{L \sigma}(E)$ describe incoming and outgoing particles with fixed energy and spin.\\
Similarly, in the topological side of the system, we obtain:
\begin{eqnarray}
\label{eq:fieldTi}
\hat{\psi}_T(x,t)&=&\int dE \Big ( \chi_{R\uparrow}(x,t)|1\rangle+\chi_{L\downarrow}(x,t)|2\rangle\Big)\nonumber\\
\hat{\psi}_B(x,t)&=&\int dE \Big ( \chi_{R\downarrow}(x,t)|3\rangle+\chi_{L\uparrow}(x,t)|4\rangle \Big),
\end{eqnarray}
where the notation
\begin{eqnarray}
\label{eq:fieldTi2}
&&\chi_{R\sigma}(x,t)=\frac{e^{-iEt/\hbar}}{\sqrt{2 \pi \hbar v}}e^{ik_{E}x}\hat{b}_{R \sigma}(E)\\
\label{eq:fieldTi3}
&&\chi_{L\sigma}(x,t)=\frac{e^{-iEt/\hbar}}{\sqrt{2 \pi \hbar v}}e^{-ik_{E}x}\hat{a}_{L \sigma}(E)\\
&&k_{E}=\frac{E}{\hbar v}>0
\end{eqnarray}
has been introduced. In writing Eq. (\ref{eq:fieldTi}), we have taken into account that states with positive helicity, represented by $\hat{\psi}_T(x,t)$, are spatially separated from states with negative helicity, described by $\hat{\psi}_B(x,t)$.\\
Once the scattering fields are known, current density operators (in units of the electron charge $q=-e$, $e>0$) can be written in the following form:
\begin{eqnarray}
\label{eq:chop}
&& \hat{J}^{c}_F=\sum_{\sigma}\int \frac{dE dE'}{2 \pi \hbar}e^{-it(E'-E)/\hbar}\nonumber\\
&& \times \Bigl [\hat{a}^{\dag}_{R \sigma}(E)\hat{a}_{R \sigma}(E')-\hat{b}^{\dag}_{L \sigma}(E)\hat{b}_{L \sigma}(E') \Bigl]\\
&& \hat{J}^{c}_T=\int \frac{dE dE'}{2 \pi \hbar}e^{-it(E'-E)/\hbar}\nonumber\\
&& \times \Bigl [\hat{b}^{\dag}_{R \uparrow}(E)\hat{b}_{R \uparrow}(E')-\hat{a}^{\dag}_{L \downarrow}(E)\hat{a}_{L \downarrow}(E') \Bigl]\\
&& \hat{J}^{c}_B=\int \frac{dE dE'}{2 \pi \hbar}e^{-it(E'-E)/\hbar}\nonumber\\
&& \times \Bigl [\hat{b}^{\dag}_{R \downarrow}(E)\hat{b}_{R \downarrow}(E')-\hat{a}^{\dag}_{L \uparrow}(E)\hat{a}_{L \uparrow}(E') \Bigl].
\end{eqnarray}
Due to the spin imbalance in the ferromagnetic electrode, a spin current described by the operator (in units of $\hbar/2$)
\begin{eqnarray}
\label{eq:Sop}
&& \hat{J}^{s}_F=\sum_{\sigma} \sigma \int \frac{dE dE'}{2 \pi \hbar}e^{-it(E'-E)/\hbar}\nonumber\\
&& \times \Bigl [\hat{a}^{\dag}_{R \sigma}(E)\hat{a}_{R \sigma}(E')-\hat{b}^{\dag}_{L \sigma}(E)\hat{b}_{L \sigma}(E') \Bigl]
\end{eqnarray}
is also generated when a voltage bias is applied to the system. The observables of the theory are the quantum statistical averages $\langle\hat{J}^{c}_F\rangle$, $\langle\hat{J}^{c}_T\rangle$, $\langle\hat{J}^{c}_B\rangle$ and $\langle\hat{J}^{s}_F\rangle$, which can be computed by explicitly using the scattering relation $\hat{b}_{j}(E)=\sum_{i} S_{ji}(E)\hat{a}_i(E)$ complemented by the electrodes correlations $\langle \hat{a}^{\dag}_j(E)\hat{a}_i(E')\rangle=\delta_{ij}\delta(E-E')f_{i}(E)$. In writing the correlation functions, the following shortened notation has been introduced: $\hat{b}_1\equiv \hat{b}_{L\uparrow}$, $\hat{b}_2\equiv \hat{b}_{L\downarrow}$, $\hat{b}_3\equiv \hat{b}_{R\uparrow}$, $\hat{b}_4\equiv \hat{b}_{R\downarrow}$, $\hat{a}_1\equiv \hat{a}_{R\uparrow}$, $\hat{a}_2\equiv \hat{a}_{R\downarrow}$, $\hat{a}_3\equiv \hat{a}_{L\downarrow}$, $\hat{a}_4\equiv \hat{a}_{L\uparrow}$ (see Fig. \ref{figure2}). Moreover, due to the system geometry, the Fermi distributions of the electrodes are given by $f_{1}(E)=f_{2}(E)=f_L(E)$ and $f_{3}(E)=f_{4}(E)=f_R(E)$.\\
Proceeding as detailed above, the charge currents expectation values take the form:
\begin{eqnarray}
\label{eq:expectationCh}
&& J^{c}_F=\sum_{j \in \{1,2\}}\int \frac{dE}{2 \pi \hbar} \Bigl [ f_{j}(E)-\sum_{r}|S_{jr}(E)|^2f_r(E)\Bigl]\nonumber\\
&& J^{c}_T=\int \frac{dE}{2 \pi \hbar}\Bigl [ \sum_{r}|S_{3r}(E)|^2 f_r(E)-f_{3}(E)\Bigl]\\
&& J^{c}_B=\int \frac{dE}{2 \pi \hbar}\Bigl [ \sum_{r}|S_{4r}(E)|^2 f_r(E)-f_{4}(E)\Bigl].\nonumber
\end{eqnarray}
Due to the scattering matrix properties, in equilibrium conditions (i.e., $f_{i}(E)=f(E)$, with $i \in \{1,...,4\}$) no charge current can flow through the system and thus $J^{c}_F=J^{c}_T=J^{c}_B=0$. Moreover, being the scattering matrix a unitary operator ($S^{\dag}S=SS^{\dag}=\mathbb{I}$), current conservation in the form $J^{c}_F=J^{c}_T+J^{c}_B$ can be easily proven.\\
Similar considerations lead to the expectation value of the spin current operator, which can be written in the following form:
\begin{eqnarray}
\label{eq:expectationSp}
&& J^{s}_F=\sum_{j \in \{1,2\}}\sigma_{j}\int \frac{dE}{2 \pi \hbar} \Bigl [ f_{j}(E)-\sum_{r}|S_{jr}(E)|^2f_r(E)\Bigl]\nonumber\\
\end{eqnarray}
with $\sigma_{1}=1$ and $\sigma_{2}=-1$. Charge and spin currents depend on the scattering matrix elements. The scattering matrix, on its turn, can be obtained from the $\mathcal{M}$-matrix as explained in Appendix \ref{sec:MtoS}.

\subsection{Boundary conditions of the scattering problem and derivation of the $\mathcal{M}$-matrix}
The presence of a velocity gradient at the interface induces non-trivial boundary conditions. The latter are the manifestation of the charge current conservation which provides a constraint for the matching matrix. We briefly discuss this topic. The charge continuity equation allows us to write the relations $J^{c}_{F}=\Psi^{\dag}(0^-)\check{v}_L\Psi(0^-)$ and $J^{c}_{T}+J^{c}_{B}=\Psi^{\dag}(0^+)\check{v}_R\Psi(0^+)$ with $J^{c}_{F}=J^{c}_{T}+J^{c}_{B}$. Due to the presence of a velocity gradient at the interface, the first quantization velocity operator on the left side of the junction is given by
\begin{eqnarray}
\check{v}_L=\left[
              \begin{array}{cccc}
                v^{+}_{F} & 0 & 0 & 0\\
                0 & -v^{-}_{F} & 0 & 0 \\
                0 & 0 & v^{-}_{F} & 0 \\
                 0 & 0 & 0 & -v^{+}_{F} \\
              \end{array}
            \right],
\end{eqnarray}
while on the right side of the junction the analogous operator takes the form:
\begin{eqnarray}
\check{v}_R=\left[
              \begin{array}{cccc}
                v & 0 & 0 & 0\\
                0 & -v & 0 & 0 \\
                0 & 0 & v & 0 \\
                 0 & 0 & 0 & -v \\
              \end{array}
            \right].
\end{eqnarray}
The current conservation and the matching condition expressed by Eq. (\ref{eq:matching}) lead to the important relation
\begin{equation}
\label{eq:interfaceCurrcons}
\mathcal{M}^{\dag} \check{v}_R \mathcal{M}=\check{v}_L,
\end{equation}
which immediately implies that the simple boundary condition $\Psi(0^+)=\Psi(0^-)$ is not allowed in the presence of a velocity gradient at the interface.
\begin{figure}[!h]
\includegraphics[scale=0.55]{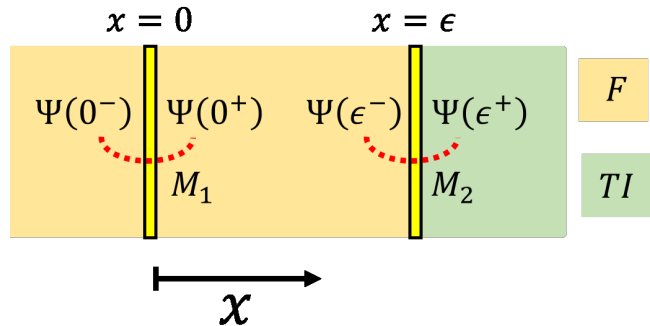}
\caption{Interface model explained in the main text. An interface potential proportional to the Dirac delta function is introduced at $x=0$, while a velocity gradient is present at $x=\epsilon$. When the limit $\epsilon \rightarrow 0^+$ is considered, the matching matrix of the resulting interface $\mathcal{M}=M_2 M_1$ is recognized.}
\label{figure3}
\end{figure}
Indeed, in the absence of interface potentials, we can look for a solution of the matrix equation in Eq. (\ref{eq:interfaceCurrcons}) in the form of a diagonal matrix. Direct computation shows that the required solution is given by:
\begin{eqnarray}
M_{2}=\mathrm{diag}\Bigl (\sqrt{\frac{v^{+}_F}{v}},\sqrt{\frac{v^{-}_F}{v}},\sqrt{\frac{v^{-}_F}{v}},\sqrt{\frac{v^{+}_F}{v}} \Bigl),
\end{eqnarray}
which coincides with the identity matrix in the absence of velocity gradient (i.e., for $ v^{\pm}_F \rightarrow v $).\\
We are now ready to treat the problem of the interface potential generated by the tunneling Hamiltonian $H_t$. Let us assume that the tunneling Hamiltonian generates an interface potential proportional to the Dirac delta function $\delta(x)$ centered at the interface point. This assumption, however, is problematic because the coexistence of the Dirac delta potential with a velocity gradient introduces analytic difficulties in the theory. The problem can be solved by introducing an infinitesimal shift $\epsilon>0$ between the point at which the interface potential diverges and the one at which the velocity gradient is present. In practice, we introduce an interface potential proportional to the Dirac delta function at $x=0$, while we set the point at which a velocity gradient is present at $x=\epsilon$, where the limit $\epsilon \rightarrow 0^+$ is implied (see Fig. \ref{figure3}). Once the diverging potential has been specified, its matching matrix $M_1$ can be computed. On the other hand, the matching matrix at the velocity gradient interface coincides with $M_2$ \cite{BCs}. Thus, the matching conditions at the two distinct interfaces can be written as follows:
\begin{eqnarray}
\Psi(0^+)=M_1 \Psi(0^-)\nonumber\\
\Psi(\epsilon^+)=M_2 \Psi(\epsilon^-).
\end{eqnarray}
In the limit $\epsilon \rightarrow 0^+$, the approximate equality $\Psi(\epsilon^-) \approx \Psi(0^+)$ becomes an exact relation and thus we get
\begin{equation}
\Psi(\epsilon^+)=M_2 M_1 \Psi(0^-).
\end{equation}
Thus, in the limit of coalescing interfaces the matching matrix of the single-interface problem can be written in the factorized form $\mathcal{M}=M_2 M_1$. The remaining part of the present subsection is devoted to the identification of $M_1$.\\
Let us add to the ferromagnet Hamiltonian (Eq. (\ref{eq:schrodingerF})) the scattering potential $\check{\mathcal{U}} \delta(x)$ with $\check{\mathcal{U}}$ a generic $2 \times 2$ Hermitian operator. The Dirac delta potential implies the usual boundary conditions for the wave functions of the ferromagnetic region:
\begin{eqnarray}
\label{eq:BCsF}
&& \partial_x\psi(0^+)-\partial_x\psi(0^-)=\frac{2}{\hbar^2}\left[
                                            \begin{array}{cc}
                                              m_{\uparrow} & 0 \\
                                              0 & m_{\downarrow}\\
                                            \end{array}
                                          \right]\check{\mathcal{U}}\psi(0^+)\nonumber\\
&& \psi(0^+)=\psi(0^-),
\end{eqnarray}
with $\psi(0^{\pm})=(\psi_{\uparrow}(0^{\pm}),\psi_{\downarrow}(0^{\pm}))^t$. Starting from Eq. (\ref{eq:BCsF}) and using the low-energy mapping (Eq. (\ref{eq:lrmovers})), the boundary conditions in the right and left movers representation can be obtained. In deriving the low-energy boundary conditions, the approximation $\partial_x \psi_{\sigma}(0^{\pm}) \approx i k^{\sigma}_{F}[ \psi_{R \sigma}(0^{\pm})-\psi_{L \sigma}(0^{\pm})]$ is required. Once the boundary conditions have been obtained, the matching matrix $M_1$ can be easily recognized.\\
We now specialize our reasonings to the scattering potential:
\begin{equation}
\check{\mathcal{U}}=\frac{\hbar v}{2} \left[
                      \begin{array}{cc}
                        g_0 & -g_1 e^{-i \theta} \\
                        -g_1 e^{i \theta}& g_0 \\
                      \end{array}
                    \right],
\end{equation}
where the dimensionless parameters $g_0$ and $g_1$ represent the spin-preserving and the spin-flipping scattering strength, respectively. Spin-flipping scattering events are activated by the interface magnetization which, in general, may well differ from the bulk value. To mimic this interface effect, we have introduced the $\theta$ parameter describing the orientation of the interface magnetization in the $x-y$ plane. With this choice, the required matching matrix takes the following form:
\begin{eqnarray}
M_1=\left[
      \begin{array}{cccc}
        1-i\alpha & i \beta & i \beta & -i \alpha \\
        -i \gamma & 1+ i \delta & i \delta & -i \gamma \\
        i \gamma & -i \delta & 1-i \delta & i \gamma \\
        i \alpha & -i \beta & -i \beta & 1+ i \alpha \\
      \end{array}
    \right],
\end{eqnarray}
where the dimensionless parameters $\alpha, \beta, \gamma, \delta$ depend on the microscopic details of the interface as specified below:
\begin{eqnarray}
\alpha=g_{0}\frac{v}{2 v^{+}_{F}}, \ \beta=g_{1}\frac{v}{2 v^{+}_{F}}e^{-i \theta}, \ \gamma=g_{1}\frac{v}{2 v^{-}_{F}}e^{i \theta}, \ \delta=g_{0}\frac{v}{2 v^{-}_{F}}.\nonumber
\end{eqnarray}
Once the matching matrices $M_1$ and $M_2$ are known, the scattering matrix is derived by using the method reported in Appendix \ref{sec:MtoS}. The explicit expression of the S-matrix is reported in Appendix \ref{sec:ExpS}.

\section{Results and discussion}
\label{sec:results}
In order to discuss the outcome of the proposed theory, we resort to a linear response formulation for the charge and spin currents. In particular the charge currents can be written (by restoring the charge prefactor) as follows:
\begin{eqnarray}
\label{eq:linrespchF}
&& J^{c}_F = \frac{e^2}{2 \pi \hbar}\sum_{j \in \{1,2\}}\Big[|S_{j3}|^2+|S_{j4}|^2 \Big](V_L-V_R)\\
\label{eq:linrespchT}
&& J^{c}_T = \frac{e^2}{2 \pi \hbar}\Big[|S_{31}|^2+|S_{32}|^2 \Big](V_L-V_R)\\
\label{eq:linrespchB}
&& J^{c}_B = \frac{e^2}{2 \pi \hbar}\Big[|S_{41}|^2+|S_{42}|^2 \Big](V_L-V_R),
\end{eqnarray}
where the scattering amplitudes are evaluated at the Fermi energy, while the quantity $V_L-V_R$ represents the voltage bias applied to the device. A similar expression can be obtained for the spin current in the ferromagnetic electrode:
\begin{equation}
\label{eq:linrespspF}
J^{s}_F = \frac{q(V_L-V_R)}{4 \pi}\sum_{j \in \{1,2\}}\sigma_{j}\Big[|S_{j3}|^2+|S_{j4}|^2 \Big],
\end{equation}
with $\sigma_1=1$ and $\sigma_2=-1$. Instead of working with the spin current $J^{s}_F$, a more useful concept in spintronics is the notion of contact polarization $P_c$ [\onlinecite{soulen98}]. This quantity, also known as current polarization, provides a measure of the degree of spin polarization of a current originated by a magnetic region. Contact polarization is in general different from the bulk polarization $P$ of a magnetic material, which is determined by the spin-dependent density of states. Despite this difference, the contact polarization $P_c$ represents the relevant quantity in understanding spin-polarized transport phenomena in nanostructured devices. Contact polarization is experimentally accessible and can be studied within the scattering theory. Using Eq. (\ref{eq:linrespchF}) and Eq. (\ref{eq:linrespspF}), $P_{c}$ takes the suggestive form:
\begin{equation}
\label{eq:contactpol}
P_{c}=-\frac{2\pi}{\Phi_0}\frac{J^{s}_F}{J^{c}_F}=\frac{\sum_{j \in \{1,2\}}\sigma_{j}\Big[|S_{j3}|^2+|S_{j4}|^2 \Big]}{\sum_{j \in \{1,2\}}\Big[|S_{j3}|^2+|S_{j4}|^2 \Big]}
\end{equation}
with $\Phi_0=h/2e$ the magnetic flux quantum. A close look at contact polarization formula given in Eq. (\ref{eq:contactpol}) shows that this quantity is simultaneously affected by the band structure of the ferromagnet and by the ferromagnet/topological insulator interface. Naturally, $P_c=0$ when a non-magnetic electrode is considered.\\
The electrical response of the device is completely characterized by the following relations:
\begin{eqnarray}
\label{eq:divider1}
&& J^{c}_{T}=\Bigl (\frac{1+P_c}{2} \Bigl ) J^{c}_F\\
\label{eq:divider2}
&& J^{c}_{B}=\Bigl (\frac{1-P_c}{2}\Bigl ) J^{c}_F
\end{eqnarray}
which can be proven by using Eqs. (\ref{eq:linrespchF})-(\ref{eq:linrespchB}) and observing that $S_{31}=S_{14}$, $S_{32}=S_{13}$, $S_{41}=S_{24}$ and $S_{42}=S_{23}$ (see Appendix \ref{sec:ExpS}). Equations (\ref{eq:divider1})-(\ref{eq:divider2}) are the main result of this work and constitute the working principle of a topological current divider. In this device, the polarized current $J^{c}_{F}$ generated by the ferromagnetic electrode is asymmetrically partitioned in spatially separated branch currents, namely $J^{c}_{T}$ and $J^{c}_{B}$, sustained by quantum states with opposite helicity. Differently from a common current divider whose response is determined by the electrical resistances of the branches, here the relative intensity of the currents in the topological side of the system is controlled by the polarization degree of the current entering the topological insulator. Along with the interest in spintronics, the mentioned effect seems to be promising in achieving an experimental estimation of $P_c$ starting from a direct measurement of $J^{c}_{T}$ and $J^{c}_{B}$ according to the relation:
\begin{equation}
P_{c}=\frac{J^{c}_{T}-J^{c}_{B}}{J^{c}_{T}+J^{c}_{B}}.
\end{equation}
The experimental effectiveness of this procedure can be altered by the unintentional introduction during the fabrication process of a resistive asymmetry between the top and bottom branches. This effect is typically due to the contact resistance formed at the interface between the topological insulator and the metallic electrodes. This problem can be mitigated by an accurate device design or by appropriate calibration of the device response.\\
Another important question to be answered is whether the difference between $J^{c}_{T}$ and $J^{c}_{B}$ is sizeable and can be easily measured when realistic $P_c$ values are considered. Certainly, a positive answer to this question depends on the microscopic details of the ferromagnetic electrode and on the characteristics of the ferromagnet/topological insulator interface. For this reason, in the following, we provide a careful discussion of this point. The contact polarization can be analytically evaluated and takes the following form:
\begin{widetext}
\begin{equation}
\label{eq:pcont}
P_c=\frac{g^{2}_0[m_{-}-m_{+}+(m_{-}+m_{+})\chi]}{g^{2}_0[m_{-}+m_{+}+(m_{-}-m_{+})\chi]+2 g^{2}_1\sqrt{m_{+}m_{-}(1-\chi^2)}+8 \eta^2 (1-\chi^2)},
\end{equation}
\end{widetext}
where the dimensionless quantities $m_{\pm}=m_{\uparrow,\downarrow}/m$, $\chi=h_{ex}/E_F$ and $\eta=\sqrt{2E_F/(m v^2)}$ have been introduced along with the bare electron mass $m$. Equation (\ref{eq:pcont}) shows that $P_c\rightarrow 0$ when $g_0\rightarrow 0$, the latter regime corresponding to a reflectionless contact between the ferromagnetic electrode and the topological insulator. Due to orbital reconstruction phenomena at the interface, the reflectionless regime mentioned above is not expected to be established in real devices and, for this reason, a correct modeling of the interface requires $g_0 \neq 0$. In the more realistic tunnel regime ($g_0\gg 1$), $P_c$ takes sizeable values which are only affected by the microscopic details of the ferromagnetic lead without any reference to the interface properties. Under this regime, it is possible to demonstrate that $J^{c}_F \propto (v^{+}_F)^2+(v^{-}_F)^2$ and $J^{s}_F \propto (v^{+}_F)^2-(v^{-}_F)^2$ so that
\begin{eqnarray}
P_{c} \approx \frac{(v^{+}_F)^2-(v^{-}_F)^2}{(v^{+}_F)^2+(v^{-}_F)^2}\equiv P^{\ast}_c,
\end{eqnarray}
the latter being in qualitative agreement with Ref. [\onlinecite{mazin}]. It is interesting to compare the above result with the bulk polarization $P$ expected for a one-dimensional electrode. The spin-dependent density of states in one dimension is given by $n_{\sigma} \propto 1/v^{\sigma}_F$ so that
\begin{eqnarray}
P & = &\frac{n_{+}-n_{-}}{n_{+}+n_{-}}=\frac{v^{-}_F-v^{+}_F}{v^{+}_F+v^{-}_F}=\nonumber\\
& = &-\frac{(v^{+}_F)^2-(v^{-}_F)^2}{(v^{+}_F)^2+(v^{-}_F)^2+2 v^{+}_F v^{-}_F}.
\end{eqnarray}
In view of the above results, we conclude that $P$ is in general different from $P^{\ast}_c$ and, as also noticed in Ref. [\onlinecite{tatsang}], a measure of $P_c$ cannot be directly used to extract information about $P$. Despite this general statement, we do observe that when the velocity range $v^{+}_F/v^{-}_F \in[0.65, 1.45]$ is considered, i.e. for $|P|\leq 0.2$, the expression $P \approx -P^{\ast}_c/2$ is very well verified. Furthermore, considering the extended velocity range $v^{+}_F/v^{-}_F \in[0.3, 3]$ ($|P|\leq 0.5$), the less accurate formula $P \approx -0.6 P^{\ast}_c$ is found. Interestingly, the sign of $P$ does not coincide with that of $P^{\ast}_c$, the latter finding being consistent with Ref. [\onlinecite{nadgorny}].\\
\begin{figure}[!h]
\includegraphics[scale=0.825]{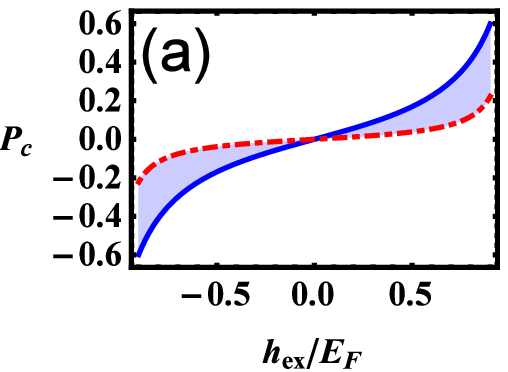}
\includegraphics[scale=0.825]{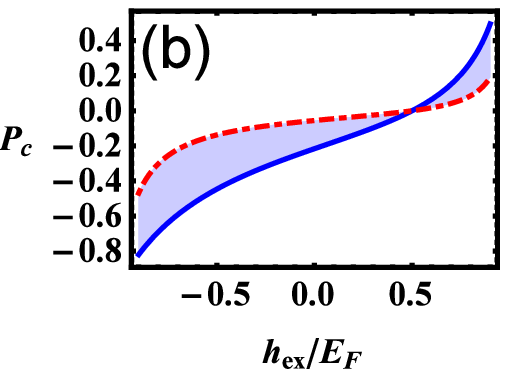}\\
\includegraphics[scale=1.2]{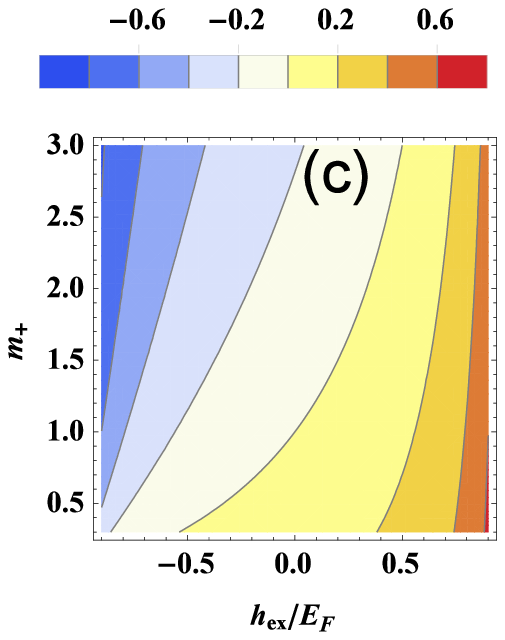}
\caption{Panel (a): $P_c$ as a function of $h_{ex}/E_F$ obtained by setting the model parameter as $m_{\pm}=1$, $g_0=2.5$, $g_1=0.5$, $\eta=5$ (dashed line) or $\eta=2$ (full line). Panel (b): $P_c$ as a function of $h_{ex}/E_F$ obtained by setting $m_{+}=3$ and $m_{-}=1$, while taking the remaining parameters as in panel (a). Panel (c): Densityplot of $P_c$ as a function of $h_{ex}/E_F$ and $m_{+}$. The remaining model parameters have been fixed as $m_{-}=1$, $g_0=2.5$, $g_1=0.5$, $\eta=2$.}
\label{figure4}
\end{figure}
Once a connection between $P$ and $P^{\ast}_c$ has been established, we provide a numerical estimation of $P_c$ based on Equation (\ref{eq:pcont}). Before studying $P_c$, we have to define the range of variability of the dimensionless parameter $\eta$ and its physical meaning. To this purpose, let us write the parameter in the evocative form $\eta=v_{\ast}/v$ with $v_{\ast}=\sqrt{2 E_F/m}$ a characteristic velocity of magnitude comparable to the Fermi velocity in a metal. Based on this observation, $v_{\ast} \sim (1 \div 2) \cdot 10^{6}$ m/s, while the typical velocity associated to the edge modes of a two-dimensional topological insulator is given by $v \sim (3.8 \div 5) \cdot 10^5$ m/s [\onlinecite{dolcini1}]. From these estimates, we conclude that $2 \lesssim \eta \lesssim 5$. When the Fermi energy can be effectively tuned by using an electrostatic back gate, the device working point can be changed and this change is accounted by a different value of $\eta$.\\
The general behavior of the contact polarization is reported in Figure \ref{figure4}. In particular, in Figure \ref{figure4} (a) the $P_c$ \textit{vs} $h_{ex}/E_F$ curve of a Stoner ferromagnet ($m_{\pm}=1$) is considered. The interface parameters have been fixed to $g_0=2.5$ and $g_1=0.5$ which correspond to an almost metallic contact with reflection probability $ \sim 20 \div 30 \%$. The contact polarization curve is a growing function of the exchange term $h_{ex}/E_F$ which presents a sensitive dependence on the $\eta$ parameter. For $\eta=2$ and $h_{ex}/E_F=0.5$, we obtain $P_c \approx 0.2$ and consequently $J^{c}_T \approx 1.5 \ J^{c}_B$. The above observation shows that the proposed device reacts to the injection of a moderate spin polarized current with a sizeable difference between $J^{c}_T$ and $J^{c}_B$.\\
When a mass asymmetry between the charge carriers with opposite spin projection is considered, the curves shown in panel (a) of Figure \ref{figure4} are deformed. This behavior is shown in Figure \ref{figure4} (b) where the $P_c$ \textit{vs} $h_{ex}/E_F$ curves are presented by setting $m_{+}=3$ and $m_{-}=1$, while taking the remaining parameters as in panel (a). Let us consider the case $h_{ex}/E_F=0$ which is appropriate to describe a ferromagnetic state only induced by a mass asymmetry. Under this circumstance, considering $\eta=2$, we get $P_c \approx -0.3$ which implies the relation $J^{c}_B \approx 1.86 \ J^{c}_T$.\\
The behavior of the contact polarization as a function of $h_{ex}/E_F$ and $m_{+}$ is presented in Figure \ref{figure4} (c). Regions with positive and negative $P_c$ values are separated by the curve $[m_{-}-m_{+}+(m_{-}+m_{+})\chi]=0$, while a prevalence of negative $P_c$ values is observed.\\
A relevant figure of merit of the topological current divider is the current ratio $J^{c}_T/J^{c}_B$ which is studied in Figure \ref{figure5}.
\begin{figure}[!h]
\includegraphics[scale=1.5]{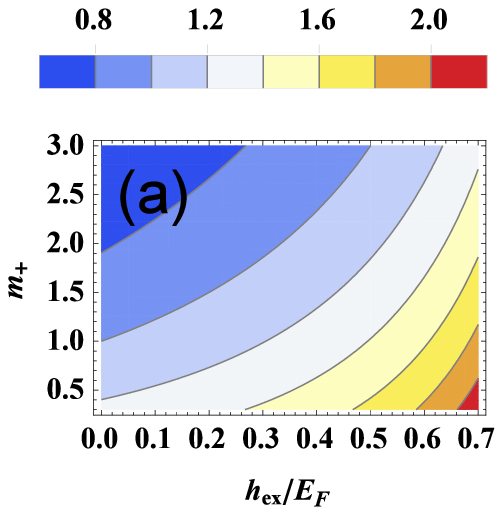}\\
\includegraphics[scale=1.5]{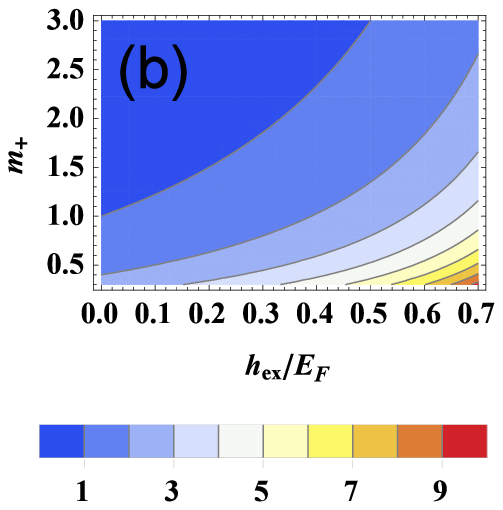}
\caption{Panel (a): $J^{c}_T/J^{c}_B$ as a function of $h_{ex}/E_F$  and $m_{+}$ obtained by setting the model parameter as $m_{-}=1$, $g_0=2.5$, $g_1=0.5$, $\eta=2$ (metallic regime). Panel (b): $J^{c}_T/J^{c}_B$ as a function of $h_{ex}/E_F$  and $m_{+}$ obtained by setting the model parameter as $m_{-}=1$, $g_0=9$, $g_1=0.5$, $\eta=2$ (tunneling regime).}
\label{figure5}
\end{figure}
In Figure \ref{figure5} (a) the current ratio $J^{c}_T/J^{c}_B$ as a function of $h_{ex}/E_F$ and $m_{+}$ is studied. Direct inspection of the figure, which is obtained within the metallic regime of the junction, shows that relevant values of the current asymmetry (i.e. $J^{c}_T/J^{c}_B \gtrsim 1.2$) can be reached even when moderate intensities (i.e. $h_{ex}/E_F\gtrsim 0.4$ with $m_{\pm}=1$) of the exchange term are considered. The tunnel limit is analyzed in Figure \ref{figure5} (b). Tunneling limit is obtained by setting the interface scattering strength to $g_0=9$, while maintaining the remaining parameters as fixed in panel (a). In this way an opaque interface with reflection probability of $\sim 80 \%$ is obtained. The analysis of the device response in tunneling regime evidences a relevant enhancement of the current asymmetry. In particular, when the parameters $h_{ex}/E_F \sim 0.45$ and $m_{\pm}=1$ are considered, we get $J^{c}_T/J^{c}_B \approx 2$.\\
The above reasonings, show that, even without considering half-metallic electrodes ($P=1$), the proposed device is able to generate a relevant current asymmetry between $J^{c}_T$ and $J^{c}_B$, which is a crucial requisite for the experimental verification. The mentioned effect is present both in tunneling and in metallic regime of the ferromagnet/topological insulator interface and presents a dependence on the microscopic details of the ferromagnet band structure. The latter dependence can be exploited to investigate conventional or exotic ferromagnetic materials, the former being described by the Stoner theory while, the ferromagnetic state of the latter is triggered by a spin-sensitive mass renormalization.\\
The proposed device can be realized by lateral etching of a quantum spin Hall material and successive deposition of a quasi-one-dimensional ferromagnetic lead to form the geometry given in Figure \ref{figure1}. In principle, before the deposition of the ferromagnetic electrode, an oxide layer (e.g. aluminium oxide of nanometer thickness) can be deposited to form a tunnel barrier between the ferromagnetic electrode and the topological insulator. In this way, the ferromagnetism acts perturbatively on the topological side of the device allowing an equilibrate interplay between competing states of matter. From the technological side, the proposed device can be fabricated by adapting the experimental process described in Refs. [\onlinecite{tianTI3d}, \onlinecite{li}]. When the quantum spin Hall state is implemented by using HgTe quantum wells, well-established multiterminal transport techniques are available at ultra-low temperature ($\sim 30$ mK) [\onlinecite{qsh-multi}]. Under this condition, the device operation in ballistic regime can be analyzed. In order to guarantee ballistic transport, a source-drain distance not-exceeding $\sim 1 \ \mu$m is necessary. Compared to the interferometric devices based on helical edge modes\cite{dolcini1}, we expect that the proposed device is much more robust against coherence losses originated by inelastic scattering events. Indeed, the topological current divider works because of the spin-momentum locking which is maintained as long as the system is described by the quantum spin Hall state. This state of matter has been proven to be robust against phonon-induced backscattering and Coulomb interaction effects\cite{dolcini2}. Robustness against large magnetic fields has been also demonstrated\cite{expQSHB,thQSHB}.

\section{Conclusions}
\label{sec:concl}
We have presented the minimal model of a ferromagnet/topological insulator device in ballistic regime. In this device, the massive states of the ferromagnetic electrode are hybridized with the helical modes of a two-dimensional topological insulator presenting the spin-momentum locking property. Resorting to a one-dimensional effective model, we have analyzed the the non-trivial boundary conditions at the ferromagnet/topological insulator interface and we have studied the device response by using the scattering field theory \`{a} la B\"{u}ttiker. We have demonstrated that the polarized current originated by the ferromagnetic electrode is asymmetrically partitioned in two branch currents sustained by quantum states with opposite helicity and belonging to the top and bottom edge of the topological insulator. The spatial separation of the branch currents allows direct measurement of the effect. The analytic evaluation of the branch currents shows that they depend on the current polarization in a simple form which can be used to measure the polarizing effect of the ferromagnetic lead. With this purpose, we have studied the contact polarization as a function of the details of the band structure of the ferromagnetic electrode. The results of this analysis suggest that the presented method could be a useful tool to measure the current polarization. When the main figure of merit of the device is analyzed, namely the current ratio $J^{c}_T/J^{c}_B$, we have found that both for metallic and tunneling contact the quantity $J^{c}_T/J^{c}_B$ is sensibly different from $1$ as long as the contact polarization is different from zero. The mentioned effect represents the working principle of a topological current divider which is of utmost interest in spintronics. Apart from its intrinsic interest for spintronics, the mentioned effect could be also important for the correct operation of devices based on topological superconductivity and Majorana fermions physics. A topological superconductor can be obtained by proximization of the edge states of a two-dimensional topological insulator with a conventional superconductor. In complex devices magnetic electrodes can be also present. We argue that under this condition the effect we are reporting can play some role in affecting the device response.\\
Finally, we have presented a minimal model serving as proof-of-principle for the operation of a topological current divider. Material-dependent information and \textit{ab initio} studies are required to get accurate estimates about the mentioned effect and eventually extend the operation conditions up to room temperature.

\appendix

\section{Derivation of the S-matrix from the $\mathcal{M}$-matrix}
\label{sec:MtoS}
Let us assume that the $\mathcal{M}$-matrix is known. The scattering fields at fixed energy $E$ in close vicinity of the interface are related as specified in Eq. (\ref{eq:matching}) of the main text. Using the scattering fields definitions given in Eqs. (\ref{eq:fieldF2})-(\ref{eq:fieldF3}) and Eqs. (\ref{eq:fieldTi2})-(\ref{eq:fieldTi3}), we can write:
\begin{eqnarray}
\label{eq:psiTiapp}
\Psi(0^+)=\left[
            \begin{array}{c}
              \chi_{R\uparrow}(0^+,t) \\
              \chi_{L\downarrow}(0^+,t) \\
              \chi_{R\downarrow}(0^+,t) \\
              \chi_{L\uparrow}(0^+,t) \\
            \end{array}
          \right]\propto \frac{1}{\sqrt{v}}\left[
                                             \begin{array}{c}
                                               \hat{b}_3(E) \\
                                               \hat{a}_3(E) \\
                                               \hat{b}_4(E) \\
                                               \hat{a}_4(E) \\
                                             \end{array}
                                           \right]
\end{eqnarray}
and
\begin{eqnarray}
\label{eq:psiFapp}
\Psi(0^-)=\left[
            \begin{array}{c}
              \phi_{R\uparrow}(0^+,t) \\
              \phi_{L\downarrow}(0^+,t) \\
              \phi_{R\downarrow}(0^+,t) \\
              \phi_{L\uparrow}(0^+,t) \\
            \end{array}
          \right]\propto \left[
           \begin{array}{c}
            \frac{\hat{a}_1(E)}{\sqrt{v^{+}_F}} \\
            \frac{\hat{b}_2(E)}{\sqrt{v^{-}_F}} \\
            \frac{\hat{a}_2(E)}{\sqrt{v^{-}_F}} \\
            \frac{\hat{b}_1(E)}{\sqrt{v^{+}_F}} \\
            \end{array}
            \right].
\end{eqnarray}
Substituting Eqs. (\ref{eq:psiTiapp})-(\ref{eq:psiFapp}) in Eq. (\ref{eq:matching}) and disregarding a common time-dependent prefactor, we get the following relation:
\begin{eqnarray}
\label{eq:matapp}
         \left[
         \begin{array}{c}
         \hat{b}_3(E) \\
         \hat{a}_3(E) \\
          \hat{b}_4(E) \\
          \hat{a}_4(E) \\
          \end{array}
          \right] = \widetilde{\mathcal{M}}
          \left[
           \begin{array}{c}
            \hat{a}_1(E) \\
            \hat{b}_2(E) \\
            \hat{a}_2(E) \\
            \hat{b}_1(E) \\
            \end{array}
            \right],
\end{eqnarray}
which is written in terms of the matrix:
\begin{eqnarray}
\widetilde{\mathcal{M}}=\mathcal{M}\cdot\mathrm{diag}\Bigl(\sqrt{\frac{v}{v^{+}_F}},\sqrt{\frac{v}{v^{-}_F}},\sqrt{\frac{v}{v^{-}_F}},\sqrt{\frac{v}{v^{+}_F}}\Bigl).
\end{eqnarray}
Eq. (\ref{eq:matapp}) can be recast in the form $\hat{b}_j=\sum_i S_{ji} \hat{a}_i$ by using the auxiliary matrices:
\begin{eqnarray}
&&\mathcal{A}=\left[
                \begin{array}{cccc}
                  0 & 0 & 1 & 0 \\
                  0 & 0 & 0 & 0 \\
                  0 & 0 & 0 & 1 \\
                  0 & 0 & 0 & 0 \\
                \end{array}
              \right],\nonumber
\mathcal{B}=\left[
                \begin{array}{cccc}
                  0 & 0 & 0 & 0 \\
                  0 & 1 & 0 & 0 \\
                  0 & 0 & 0 & 0 \\
                  1 & 0 & 0 & 0 \\
                \end{array}
              \right], \nonumber\\
&& \mathcal{C}=\left[
                \begin{array}{cccc}
                  1 & 0 & 0 & 0 \\
                  0 & 0 & 0 & 0 \\
                  0 & 1 & 0 & 0 \\
                  0 & 0 & 0 & 0 \\
                \end{array}
              \right],\nonumber
\mathcal{D}=\left[
                \begin{array}{cccc}
                  0 & 0 & 0 & 0 \\
                  0 & 0 & 1 & 0 \\
                  0 & 0 & 0 & 0 \\
                  0 & 0 & 0 & 1 \\
                \end{array}
              \right]. \nonumber
\end{eqnarray}
This procedure leads to the following relation between the S-matrix and the $\mathcal{M}$-matrix:
\begin{eqnarray}
S(E)=(\mathcal{A}-\widetilde{\mathcal{M}}\mathcal{B})^{-1}(\widetilde{\mathcal{M}}\mathcal{C}-\mathcal{D}).
\end{eqnarray}
Thus the S-matrix is determined once the $\mathcal{M}$-matrix, which specifies the ferromagnet-topological insulator coupling, is known.
\begin{widetext}
\section{Expression of the S-matrix}
\label{sec:ExpS}
The scattering matrix of the problem takes the following form:
\begin{eqnarray}
S=\left[
\begin{array}{cccc}
 \frac{-g_0^2+2 i v_{\downarrow }
   g_0+g_1^2}{g_0^2-2 i
   \left(v_{\downarrow
   }+v_{\uparrow }\right)
   g_0-g_1^2-4 v_{\downarrow }
   v_{\uparrow }} & \frac{2 i e^{-i
   \theta } g_1 \sqrt{v_{\downarrow
   } v_{\uparrow }}}{-g_0^2+2 i
   \left(v_{\downarrow
   }+v_{\uparrow }\right)
   g_0+g_1^2+4 v_{\downarrow }
   v_{\uparrow }} & \frac{2 i e^{-i
   \theta } g_1 \sqrt{v_{\downarrow
   } v_{\uparrow }}}{-g_0^2+2 i
   \left(v_{\downarrow
   }+v_{\uparrow }\right)
   g_0+g_1^2+4 v_{\downarrow }
   v_{\uparrow }} & \frac{-2 i g_0
   v_{\uparrow }-4 v_{\downarrow }
   v_{\uparrow }}{g_0^2-2 i
   \left(v_{\downarrow
   }+v_{\uparrow }\right)
   g_0-g_1^2-4 v_{\downarrow }
   v_{\uparrow }} \\
 \frac{2 i e^{i \theta } g_1
   \sqrt{v_{\downarrow }
   v_{\uparrow }}}{-g_0^2+2 i
   \left(v_{\downarrow
   }+v_{\uparrow }\right)
   g_0+g_1^2+4 v_{\downarrow }
   v_{\uparrow }} & \frac{-g_0^2+2
   i v_{\uparrow }
   g_0+g_1^2}{g_0^2-2 i
   \left(v_{\downarrow
   }+v_{\uparrow }\right)
   g_0-g_1^2-4 v_{\downarrow }
   v_{\uparrow }} & \frac{-2 i g_0 v_{\downarrow }-4 v_{\uparrow }v_{\downarrow }}{g_0^2-2 i
   \left(v_{\downarrow
   }+v_{\uparrow }\right) g_0-
   g_1^2-4 v_{\downarrow }
   v_{\uparrow }} & \frac{2
   i e^{i \theta } g_1
   \sqrt{v_{\downarrow }
   v_{\uparrow }}}{-g_0^2+2 i
   \left(v_{\downarrow
   }+v_{\uparrow }\right)
   g_0+g_1^2+4 v_{\downarrow }
   v_{\uparrow }} \\
 \frac{-2 i g_0 v_{\uparrow }-4
   v_{\downarrow } v_{\uparrow
   }}{g_0^2-2 i \left(v_{\downarrow
   }+v_{\uparrow }\right)
   g_0-g_1^2-4 v_{\downarrow }
   v_{\uparrow }} & \frac{2 i e^{-i
   \theta } g_1 \sqrt{v_{\downarrow
   } v_{\uparrow }}}{-g_0^2+2 i
   \left(v_{\downarrow
   }+v_{\uparrow }\right)
   g_0+g_1^2+4 v_{\downarrow }
   v_{\uparrow }} & \frac{2 i e^{-i
   \theta } g_1 \sqrt{v_{\downarrow
   } v_{\uparrow }}}{-g_0^2+2 i
   \left(v_{\downarrow
   }+v_{\uparrow }\right)
   g_0+g_1^2+4 v_{\downarrow }
   v_{\uparrow }} & \frac{-g_0^2+2
   i v_{\downarrow }
   g_0+g_1^2}{g_0^2-2 i
   \left(v_{\downarrow
   }+v_{\uparrow }\right)
   g_0-g_1^2-4 v_{\downarrow }
   v_{\uparrow }} \\
 \frac{2 i e^{i \theta } g_1
   \sqrt{v_{\downarrow }
   v_{\uparrow }}}{-g_0^2+2 i
   \left(v_{\downarrow
   }+v_{\uparrow }\right)
   g_0+g_1^2+4 v_{\downarrow }
   v_{\uparrow }} & \frac{-2 i g_0
   v_{\downarrow }-4 v_{\uparrow }
   v_{\downarrow }}{g_0^2-2 i
   \left(v_{\downarrow
   }+v_{\uparrow }\right)
   g_0-g_1^2-4 v_{\downarrow }
   v_{\uparrow }} & \frac{-g_0^2+2
   i v_{\uparrow }
   g_0+g_1^2}{g_0^2-2 i
   \left(v_{\downarrow
   }+v_{\uparrow }\right)
   g_0-g_1^2-4 v_{\downarrow }
   v_{\uparrow }} & \frac{2 i e^{i
   \theta } g_1 \sqrt{v_{\downarrow
   } v_{\uparrow }}}{-g_0^2+2 i
   \left(v_{\downarrow
   }+v_{\uparrow }\right)
   g_0+g_1^2+4 v_{\downarrow }
   v_{\uparrow }} \\
\end{array}
\right]\nonumber,
\end{eqnarray}
\end{widetext}
where the notation $v_{\uparrow}=v^{+}_{F}/v$ and $v_{\downarrow}=v^{-}_{F}/v$ has been introduced. Direct inspection of the S-matrix structure evidences that spin-flipping scattering events are responsible for the activation of intraedge backscattering phenomena in the topological side of the junction. Due to this, $S_{33}$ and $S_{44}$ are proportional to the spin flipping scattering strength $g_1$. These terms are vanishing quantities in the presence of preserved helicity and time reversal symmetry. On the other hand, the coupling with the ferromagnetic electrode also induces interedge scattering events in the topological side of the junction. The interedge coupling, related to $S_{34}$ and $S_{43}$, is affected by the ferromagnetism but it is not canceled when a non-magnetic electrode is considered.\\
The scattering matrix also depends on the phase factor $e^{\pm i \theta}$ which however plays no role in a single-interface device.

\section*{Acknowledgment}
Discussions with R. De Luca are gratefully acknowledged.


\begin{thebibliography}{99}
\bibitem{topo1} J. Wang, S. Zhang, Nature Materials \textbf{16}, 1062 (2017).
\bibitem{topo2} D. Castelvecchi, Nature \textbf{571}, 17 (2019).
\bibitem{topo3} M. Asorey, Nature Physics \textbf{12}, 616 (2016).
\bibitem{topo4} C. Beenakker, L. Kouwenhoven, Nature Physics \textbf{12}, 618 (2016).
\bibitem{topo5} L. Lu, J. Joannopoulos and M. Solja\v{c}i\'{c}, Nature Physics \textbf{12}, 626 (2016).
\bibitem{topo6} N. Goldman, J. Budich and P. Zoller, Nature Physics \textbf{12}, 639 (2016).
\bibitem{topo7} S. Huber, Nature Physics \textbf{12}, 621 (2016).

\bibitem{holo} L. Susskind, Journal of Mathematical Physics \textbf{36}, 6377 (1995).

\bibitem{topsuper} M. Sato and Y. Ando, Rep. Prog. Phys. \textbf{80}, 076501 (2017).
\bibitem{topins} M. Z. Hasan and C. L. Kane, Rev. Mod. Phys. \textbf{82}, 3045 (2010).
\bibitem{topsupins}B. J\"{a}ck, Y. Xie, J. Li, S. Jeon, B. A. Bernevig, A. Yazdani, Science \textbf{364}, 1255 (2019).

\bibitem{kitaev} A. Yu. Kitaev, Phys.-Uspekhi \textbf{44}, 131 (2001).
\bibitem{majoexp1} V. Mourik, K. Zuo, S. M. Frolov, S. R. Plissard, E. P. A. M. Bakkers, L. P. Kouwenhoven, Science \textbf{336}, 1003 (2012).
\bibitem{majoexp2} Stevan Nadj-Perge, Ilya K. Drozdov, Jian Li, Hua Chen, Sangjun Jeon, Jungpil Seo, Allan H. MacDonald, B. Andrei Bernevig, Ali Yazdani, Science \textbf{346}, 602 (2014).
\bibitem{theormajo} M. Sato and S. Fujimoto, J. Phys. Soc. Jpn. \textbf{85}, 072001 (2016).
\bibitem{theormajo2} J. Alicea, Rep. Prog. Phys. \textbf{75}, 076501 (2012).

\bibitem{expmajo1} S. Manna, P. Wei, Y. Xie, K. T. Law, P. A. Lee, J. S. Moodera,
Proceedings of the National Academy of Sciences \textbf{117}(16), 8775 (2020).
\bibitem{expmajo2} Hao-Hua Sun, Kai-Wen Zhang, Lun-Hui Hu, Chuang Li, Guan-Yong Wang, Hai-Yang Ma, Zhu-An Xu, Chun-Lei Gao, Dan-Dan Guan, Yao-Yi Li, Canhua Liu, Dong Qian, Yi Zhou, Liang Fu, Shao-Chun Li, Fu-Chun Zhang and Jin-Feng Jia, Phys. Rev. Lett. \textbf{116}, 257003 (2016).
\bibitem{expmajo3} Gerbold C. M\'{e}nard, Andrej Mesaros, Christophe Brun, Fran\c{c}ois Debontridder, Dimitri Roditchev, Pascal Simon and Tristan Cren,  Nat. Commun. \textbf{10}, 2587 (2019).

\bibitem{kitaev-computation} A. Yu. Kitaev, Annals of Physics \textbf{303}, 2 (2003).
\bibitem{majo-computation} H. Zhang, D. E. Liu, M. Wimmer, L. P. Kouwenhoven, Nat. Commun. \textbf{10}, 5128 (2019).

\bibitem{2dTiQW1} B. A. Bernevig, T. L. Hughes, and S.-C. Zhang, Science \textbf{314}, 1757 (2006).
\bibitem{2dTiQW2} M. K\"{o}nig, S. Wiedmann, C. Br\"{u}ne, A. Roth, H. Buhmann,
L. W. Molenkamp, X.-L. Qi, and S.-C. Zhang, Science \textbf{318}, 766 (2007).

\bibitem{3dTith} L. Fu, C. L. Kane, and E. J. Mele, Phys. Rev. lett. \textbf{98},
106803 (2007).
\bibitem{3dTi}D. Hsieh, D. Qian, L. Wray, Y. Xia, Y. S. Hor, R. J. Cava, and
M. Z. Hasan, Nature \textbf{452}, 970 (2008).


\bibitem{hoti1} W. A. Benalcazar, B. A. Bernevig, and T. L. Hughes, Science \textbf{357}, 61 (2017).
\bibitem{hoti2} W. A. Benalcazar, B. A. Bernevig, and T. L. Hughes, Phys. Rev. B \textbf{96}, 245115 (2017).
\bibitem{hoti3} J. Langbehn, Y. Peng, L. Trifunovic, F. von Oppen, and P. W. Brouwer, Phys. Rev. Lett. \textbf{119}, 246401 (2017).
\bibitem{hoti4} Z. Song, Z. Fang, and C. Fang, Phys. Rev. Lett. \textbf{119}, 246402 (2017).


\bibitem{dolcini1} Fabrizio Dolcini, Phys. Rev. B \textbf{83}, 165304 (2011).
\bibitem{dolcini2} Jan Carl Budich, Fabrizio Dolcini, Patrik Recher, and Bj\"{o}rn Trauzettel, Phys. Rev. Lett. \textbf{108}, 086602 (2012).
\bibitem{dolcini3} Pietro Sternativo and Fabrizio Dolcini, Phys. Rev. B \textbf{89}, 035415 (2014).
\bibitem{dolcini4} Fran\c{c}ois Cr\'{e}pin, Bj\"{o}rn Trauzettel, and Fabrizio Dolcini, Phys. Rev. B \textbf{89}, 205115 (2014).
\bibitem{dolcini5} Fabrizio Dolcini, Phys. Rev. B \textbf{92}, 155421 (2015).
\bibitem{dolcini6} Fabrizio Dolcini, Rita Claudia Iotti, Arianna Montorsi, and Fausto Rossi, Phys. Rev. B \textbf{94}, 165412 (2016).
\bibitem{dolcini7} Fabrizio Dolcini, Phys. Rev. B \textbf{95}, 085434 (2017).

\bibitem{romeo1} R. Citro, F. Romeo, and N. Andrei, Phys. Rev. B \textbf{84}, 161301(R) (2011).
\bibitem{romeo2} F. Romeo, R. Citro, D. Ferraro, and M. Sassetti, Phys. Rev. B \textbf{86}, 165418 (2012).
\bibitem{romeo3} D. Ferraro, G. Dolcetto, R. Citro, F. Romeo, and M. Sassetti, Phys. Rev. B \textbf{87}, 245419 (2013).
\bibitem{romeo4} F. Romeo and R. Citro, Phys. Rev. B \textbf{90}, 155408 (2014).

\bibitem{bercioux1} Andreas Inhofer and Dario Bercioux, Phys. Rev. B \textbf{88}, 235412 (2013).
\bibitem{ABferroTi} Joseph Maciejko, Eun-Ah Kim, and Xiao-Liang Qi, Phys. Rev. B \textbf{82}, 195409 (2010).
\bibitem{F-TIdev1} Yuan Li, M. B. A. Jalil, Seng Ghee Tan and GuangHui Zhou, J. Appl. Phys. \textbf{112}, 063710 (2012).
\bibitem{F-TIdev2} Junji Guo, Wenhu Liao, Heping Zhao and Guanghui Zhou, J. Appl. Phys.\textbf{115}, 023709 (2014).
\bibitem{F-TIdev3} Xianbo Xiao, Ying Liu, Zhengfang Liu, Guoping Ai, Shengyuan A. Yang and Guanghui Zhou, Appl. Phys. Lett. \textbf{108}, 032403 (2016).


\bibitem{largegapexp} R. Wu, J.-Z. Ma, S.-M. Nie, L.-X. Zhao, X. Huang, J.-X. Yin, B.-B. Fu, P. Richard, G.-F. Chen, Z. Fang, X. Dai, H.-M. Weng, T. Qian, H. Ding, and S. H. Pan, Phys. Rev. X \textbf{6}, 021017 (2016).

\bibitem{QSHEroom} Sanfeng Wu, Valla Fatemi, Quinn D. Gibson, Kenji Watanabe, Takashi Taniguchi, Robert J. Cava, Pablo Jarillo-Herrero, Science \textbf{359} (6371), 76 (2018).

\bibitem{magnetismhelicalexp} Berthold J\"{a}ck, Yonglong Xie, B. Andrei Bernevig, Ali Yazdani, PNAS \textbf{117} (28), 16214 (2020).

\bibitem{buttiker92} M. B\"{u}ttiker, Phys. Rev. B \textbf{46}, 12485 (1992).
\bibitem{blanter2000} Ya. M. Blanter and M. B\"{u}ttiker, Physics Reports \textbf{336}, 1-166 (2000).

\bibitem{chamon} Chang-Yu Hou, Eun-Ah Kim and Claudio Chamon, Phys. Rev. Lett. \textbf{102}, 076602 (2009).

\bibitem{hirsch} J. E. Hirsch, Phys. Rev. B \textbf{59}, 6256 (1999).
\bibitem{Higashiguchi} Mitsuharu Higashiguchi, Kenya Shimada, Keisuke Nishiura, Xiaoyu Cui, Hirofumi Namatame and Masaki Taniguchi, Phys. Rev. B \textbf{72}, 214438 (2005).

\bibitem{projection} Mark I. Visscher and Gerrit E. W. Bauer, Phys. Rev. B \textbf{54}, 2798 (1996).

\bibitem{BCs} The resulting boundary conditions are similar to those considered in N. M. R. Peres, Journal of Physics: Condensed Matter \textbf{21},
095501 (2009).

\bibitem{soulen98} R. J. Soulen Jr., J. M. Byers, M. S. Osofsky, B. Nadgorny, T. Ambrose, S. F. Cheng, P. R. Broussard, C. T. Tanaka, J. Nowak, J. S. Moodera, A. Barry, J. M. D. Coey, Science \textbf{282}, 85 (1998).

\bibitem{mazin} I. I. Mazin, Phys. Rev. Lett. \textbf{83}, 1427 (1998).
\bibitem{tatsang} Tat-Sang Choy, Jian Chen, and Selman Hershfield, Journal of Applied Physics \textbf{86}, 562 (1999).
\bibitem{nadgorny} B. Nadgorny, R. J. Soulen, Jr., M. S. Osofsky, I. I. Mazin, G. Laprade, R. J. M. van de Veerdonk, A. A. Smits, S. F. Cheng, E. F. Skelton, and S. B. Qadri, Phys. Rev. B \textbf{61}, R3788 (2000).

\bibitem{tianTI3d} Jifa Tian, Ireneusz Miotkowski, Seokmin Hong and Yong P. Chen, Sci. Rep. \textbf{5}, 14293 (2015).
\bibitem{li}C. H. Li, O.M.J. van't Erve, S. Rajput, L. Li and B. T. Jonker, Nat. Commun. \textbf{7}, 13518 (2016).

\bibitem{qsh-multi} Andreas Roth, Christoph Br\"{u}ne, Hartmut Buhmann, Laurens W. Molenkamp, Joseph Maciejko, Xiao-Liang Qi, Shou-Cheng Zhang, Science \textbf{325}, 294 (2009).

\bibitem{expQSHB} Lingjie Du, Ivan Knez, Gerard Sullivan, and Rui-Rui Du, Phys. Rev. Lett. \textbf{114}, 096802 (2015).
\bibitem{thQSHB} Song-Bo Zhang, Yan-Yang Zhang, and Shun-Qing Shen, Phys. Rev. B \textbf{90}, 115305 (2014).


\end{thebibliography}
\end{document}